\documentstyle [12pt] {article}
\title{PETER AND ANTI-PETER PRINCIPLE AS THE DISCRETE LOGISTIC EQUATION}

\author{Vladan Pankovi\'c\\
Department of Physics, Faculty of Sciences, 21000 Novi Sad,\\ Trg
Dositeja Obradovi\'ca 4. , Serbia, vpankovic@if.ns.ac.yu}

\date {}
\begin{document}
\maketitle
 \vspace {0.5cm}

\begin {abstract}
In this work Peter principle (in the hierarchical structure any
competent member tends to rise to his level of incompetence) is
consistently interpreted as the discrete form of the well-known
logistic (Verhulst or Maltusian) equation of the population
dynamics. According to such interpretation anti-Peter principle
(in the hierarchical structure any incompetent member tends to
rise to his level of competence) is formulated too.
\end {abstract}

As it is well-known remarkable Peter principle [1], [2] states
that in the hierarchical structure any competent member tends to
rise to his level of incompetence. Even if Peter principle is
seemingly paradoxical it is in a satisfactory agreement with
situations existing in real social hierarchical structures. There
are different attempts of the interpretation or mathematical
foundation of Peter principle. In this work an original
interpretation will be suggested. Namely, in this work Peter
principle will be consistently interpreted as the discrete
logistic (Verhulst or Maltusian) equation of the population
dynamics. According to such interpretation anti-Peter principle
(in the hierarchical structure any incompetent member tends to
rise to his level of competence) is formulated too.

Thus, as it is well-known logistic (Verhulst or Maltusian)
equation in the population dynamics has form
\begin {equation}
  \frac {dx}{dt} = {\it a}x(1-\frac {x}{r})  \hspace{1cm} {\rm for} \hspace{0.5 cm} a,r > 0 \hspace{0.5 cm}{\rm and}
  \hspace{0.5 cm} x \leq r
\end {equation}

where $t$ represents the time moment, $x$ - (human or some other
species) population, ${\it a}$  - growth parameter and r carrying
capacity. Simple solution of this equation, representing a sigmoid
function, is
\begin {equation}
 x = x_{0} r \exp[{\it a}t]\frac {1}{r - x_{0} - x_{0}\exp[{\it a}t]}
\end {equation}
where $x_{0}$ represents the initial population smaller than $r$.
Obviously, when $t$ tends toward infinity $x$ tends toward r and
$\frac {dx}{dt}$ toward zero. Given logistic dynamics describes
population growth limited by negative species self-interaction.

It is well known too that there is anti-logistic equation
corresponding to (1)
\begin {equation}
    \frac {dx}{dt}= -{\it a}x(\frac {x}{r} - 1) \hspace{1cm} {\rm for} \hspace{0.5 cm} a,r > 0 \hspace{0.5 cm}{\rm and}
                \hspace{0.5 cm} x \geq r
\end {equation}
where $-{\it a}$ represents the decrease parameter. It holds
simple solution
\begin {equation}
   x = x_{0} r \frac {1}{x_{0} - (x_{0} - r) \exp[{\it a}t]}
\end {equation}
where $x_{0}$ represents the initial population smaller than $r$.
Obviously, when $t$ tends toward infinity $x$ tends toward $r$ and
$\frac {dx}{dt}$ toward zero. Given logistic dynamics describes
population decrease limited by positive species self-interaction.

Finally, it is well-known that both, logistic and anti-logistic,
equations have significant application not only in population
dynamics, i.e. in the biology and demography, but also in the
chemistry, mathematical psychology, economics and sociology.

We observe that sigmoid form of the solution of logistic equation
satisfactory corresponds to predicted and observed form of the
successful member competence time evolution. For this reason we
suppose that, in the first approximation, successful member
competence time evolution is described by (1). But now $x$
represents competence in the time moment $t$, $x_{0}$ - initial
competence, ${\it a}$ - competence growth parameter and $r$ -
level of incompetence, all of which are characteristic for given
member.

But in distinction of a biological species where individuals
number can be very large so that population can be effectively
treated as a continuous variable satisfying logistic differential
equation (1), number of the members in a hierarchical sociological
structure (e.g. factory, university, etc.) can be relatively
small. For this reason member competence can be a discrete
function that satisfies a discrete dynamics. Nevertheless, it is
not hard to see that such discrete dynamics must correspond to the
discretized form of the logistic equation (1) (which will be not
discussed detailedly). According to such discretization any
competent member of the hierarchical structure can rise to its
incompetence level in a finite time interval.

Moreover, mentioned interpretation of the Peter principle (by
discretized form of the logistic equation) admits formulation of
the anti-Peter principle by discretized anti-logistic equation
(3). But now $x$ represents competence in the time moment $t$,
$x_{0}$ - initial competence, $-{\it a}$ - competence decrease
parameter and $r$ - level of competence, all of which are
characteristic for given member. This principle states that in the
hierarchical structure any incompetent member tends to rise to his
level of competence. It is, of course, in full agreement with
discretized form of (4) (which will be not discussed detailedly).
Also, according to such discretization, any incompetent member of
the hierarchical structure can rise to its competence level in a
finite time interval.

In conclusion, it can be shortly repeated and pointed out the
following. In this work Peter principle (in the hierarchical
structure any successful member tends to rise to his level of
incompetence) is consistently interpreted as the discrete form of
the well-known logistic (Verhulst or Maltusian) equation of the
population dynamics. According to such interpretation anti-Peter
principle (in the hierarchical structure any non-successful member
tends to rise to his level of competence) is formulated too.

\section {References}

\begin {itemize}

\item [[1]] L. J. Peter, R. Hul, {\it The Peter Principle: why things always go wrong} (William Morrow and Company, New York, 1969)
\item [[2]] A. Pulchino, A. Rapisarda, C. Garofalo, {\it The Peter Principle Revisited: A Computational Study} ,  soc-ph/0907.0455 and references therein

\end {itemize}

\end {document}